\documentclass[11pt,twoside]{article}  
\usepackage{apn3conf}

\begin{document}   

%
%
%
%

\title{HST/NICMOS Imaging Polarimetry of PPNs}

%

\author{Toshiya Ueta}
\affil{Royal Observatory of Belgium, Ringlaan 3, B-1180, Belgium}
\author{Koji Murakawa}
\affil{Subaru Telescope, 650 N. A'ohoku Place, Hilo, HI 96720, USA}
\author{Margaret Meixner}
\affil{STScI, 3700 San Martin Dr., Baltimore, MD 21218, USA}


\contact{Toshiya Ueta}
\email{ueta@oma.be}

%
%
%
%
%

\paindex{Ueta, T.}
\aindex{Murakawa, K.}     
\aindex{Meixner, M.}     

%
%

\authormark{Ueta, Murakawa, \& Meixner}

%

\keywords{IRAS 07134+1005, IRAS 06530-0213, IRAS 04296+3429, 
IRAS 02229+6208, dust, polarization, circumstellar shell}


\begin{abstract}          
We present the preliminary results from our {\sl HST}/NICMOS imaging 
polarimetry on optically thin PPNs.
The data show the dust distribution in an unprecedented detail.
The structure of these faint PPNs is revealed via polarization
properties of the dust-scattered light without much interference from
the bright central star, proving the unique advantages of the technique
over simple imaging. 
\end{abstract}


\section{Introduction}

We observed 4 PPNs of SOLE type (Ueta et al.\ 2000) by means of imaging
polarimetry
using {\sl HST}/NICMOS (NIC1 with POL0S, POL120S, and POL240S filters)
during cycle 11 (GO 9377).
Our main objective was 
to study the dust distribution in these {\sl optically thin} 
PPN shells around the bright central star.
Imaging polarimetry (e.g., Gledhill et al.\ 2001) is a very effective
technique for this purpose because 
(1) bright stellar emission (unpolarized component) is
automatically removed when the linearly polarized component 
(i.e., dust-scattered star light of the reflection nebula) is
computed from the Stokes $Q$ and $U$ parameters,
(2) the dust distribution in the reflection nebula can be probed at high
resolution as opposed to the mid-IR diffraction-limit resolution,
and
(3) dust properties (e.g., the size distribution) can also be 
investigated. 

\section{Preliminary Results}

\begin{figure}
\epsscale{1}
\plottwo{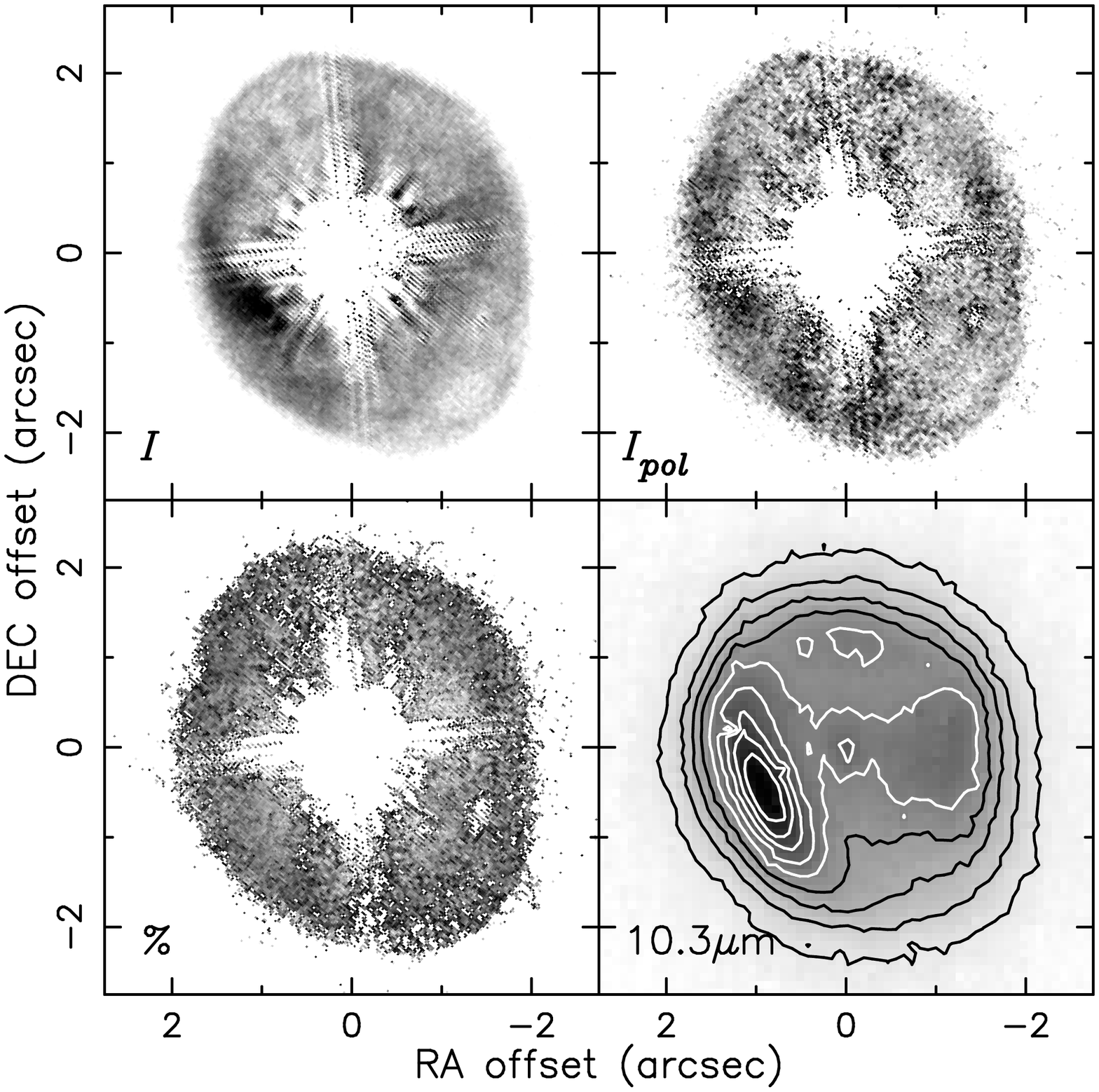}{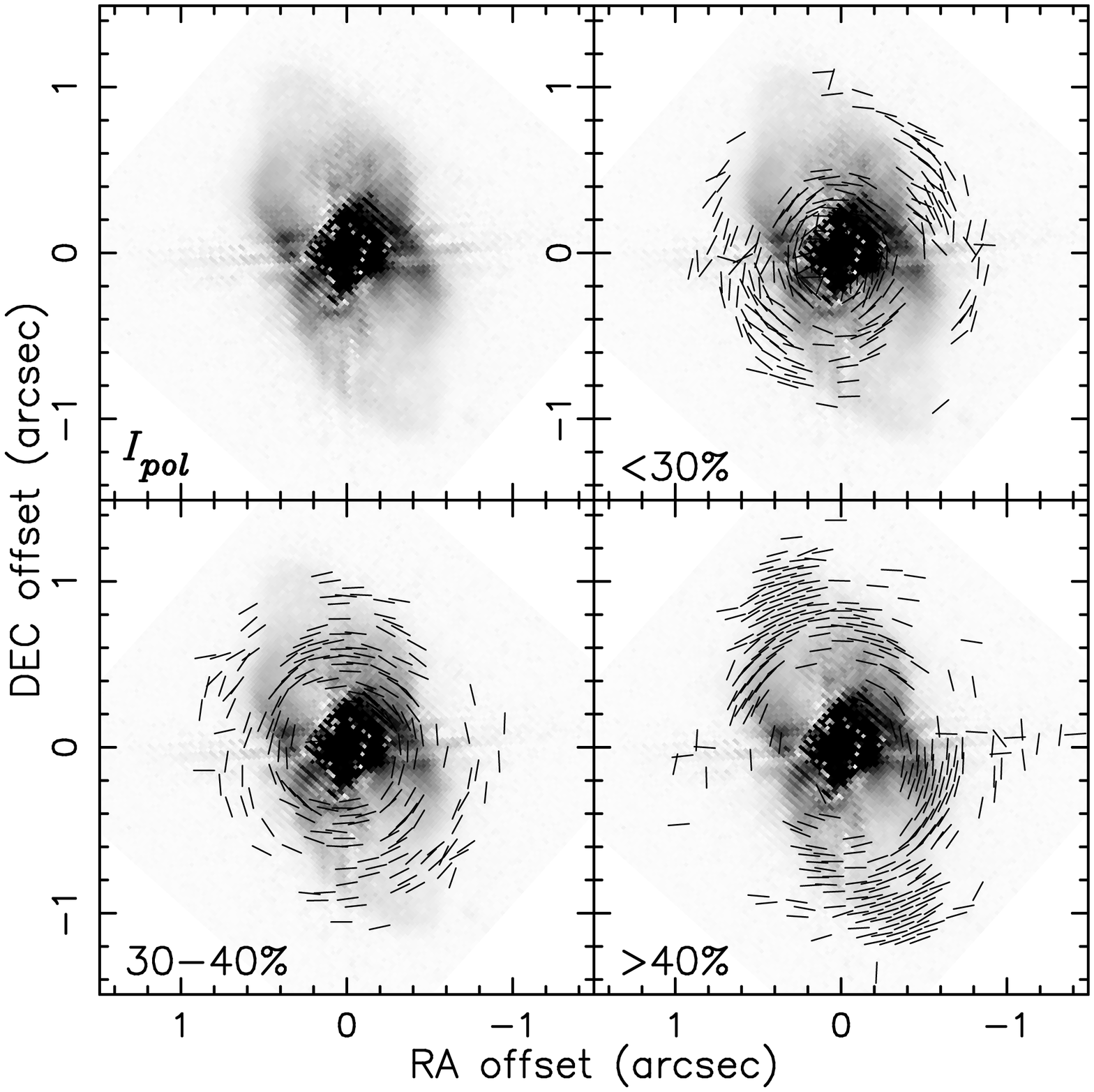}
\caption{%
Imaging polarimetry data of IRAS 07134+1005 (left) and
IRAS 06530-0123 (right).} \label{fig1}
\end{figure}

Figure \ref{fig1} shows imaging polarimetry 
data for IRAS 07134+1005 (left) and IRAS 06530-0213 (right).
The IRAS 07134+1005 data illustrate the distribution and
strength of polarization.
The four frames display, from top left to bottom right,
the total intensity (the Stokes $I$),
the polarized intensity ($I_{\rm pol} = \sqrt{Q^2 + U^2}$),
the degree of polarization ($I_{\rm pol}/I$),
and the 10.3 $\mu$m Gemini image (Kwok et al.\ 2002) 
for comparison.
The total and polarized intensity images look similar, suggesting that
the nebula emission consists mostly of scattered star light as expected
in this reflection nebula. 
The total intensity image confirms the equatorially 
enhanced (toroidal) structure of the shell as seen in the 
10.3 $\mu$m image and also shows the structure {\sl within} the 
reflection nebula in an unprecedented detail.
Such detailed structure of the shell was not seen well in 
{\sl HST}/WFPC2 data due to the invasively bright central star and its
enormous PSF spikes 
(Ueta et al.\ 2001).

In the IRAS 06530-0213 data,
the distribution and angles of polarization vectors
are displayed.
The four frames shows the the polarized intensity,
the distribution of polarization vectors of less than 30\% 
($I_{\rm pol}/I < 0.3$) strengths,
of 30 to 40\% strengths,
and of more than 40\% strengths,
from top left to bottom right.
The vectors are distributed in a clear centro-symmetric pattern
with respect to the central star. 
This confirms the optically thin nature of this PPN shell.
It is also apparent that the stronger polarization is found near the
edge of the shell.
This is because 
dust-scattered light near the nebula edge tends to have
near 90$^{\circ}$ scattering angle, and hence, tends to yield
the highest degree of polarization.
Our data yielded rather high degrees of polarization in 
our target PPNs.
This may be due to a population of small dust grains 
residing in these shells.
We are currently finalizing our data analysis, and the results would
shed light on the PPN shell structure as well as the properties of dust
grains present in PPNs.

\acknowledgments

Support for GO-9377 was provided by NASA through a grant from
the STScI, which is operated by the AURA, Inc., under NASA
contract NAS 5-26555. 
Ueta was also supported by IUAP P5/36 financed by the OSTC 
of the Belgian Federal State.

%
%
%


\end{document}